\documentclass[twocolumn,showpacs,prlprintnumbers,amsmath,amssymb, superscriptaddress]{revtex4-1}

\usepackage{graphicx}
\usepackage{amsmath,amssymb}
\usepackage{color}
\usepackage{float}
\usepackage{hyperref}

\newcommand{\ket}[1]{\ensuremath{\left| #1 \right\rangle}}
\newcommand{\bra}[1]{\ensuremath{\left\langle #1 \right|}}

\definecolor{Zcolour}{rgb}{0.992, 0.588, 0.22}

\begin{document}

\title{A Quantum Dipolar Spin Liquid}

\author{N.~Y.~Yao}
\affiliation{Physics Department,  University of California Berkeley, Berkeley, CA 94720, U.S.A.}
\author{M.~P.~Zaletel}
\affiliation{Station Q,  Microsoft Research, Santa Barbara, CA 93106, U.S.A.}
\author{D.~M.~Stamper-Kurn}
\affiliation{Physics Department,  University of California Berkeley, Berkeley, CA 94720, U.S.A.}
\author{A.~Vishwanath}
\affiliation{Physics Department,  University of California Berkeley, Berkeley, CA 94720, U.S.A.}

\begin{abstract}

Quantum spin liquids are a new class of magnetic ground state in which spins are quantum mechanically entangled over macroscopic scales. Motivated by recent advances in the control of polar molecules, we show that dipolar  interactions between $\textrm{S}=1/2$ moments stabilize spin liquids on the triangular and kagome lattices.
In the latter case, the moments spontaneously break time-reversal, forming a chiral spin liquid with robust edge modes and emergent semions.
We propose a simple route toward synthesizing a dipolar Heisenberg antiferromagnet from lattice-trapped polar molecules using only a single pair of  rotational states and a constant electric field.

\end{abstract}
\maketitle

In strongly frustrated  systems, competing interactions can conspire with quantum fluctuations to prevent classical order  down to zero temperature.
In an antiferromagnet, frustration allows magnetic moments  to evade the formation of conventional long-range order, leading  to the magnetic analog of liquid phases.
Such {\it quantum spin liquids} are characterized by entanglement over macroscopic scales and can exhibit a panoply of exotic properties, ranging from emergent gauge fields and fractionalized excitations to robust chiral edge modes \cite{Anderson1973, Kalmeyer87, Balents2010}. 
Definitively finding and characterizing such an exotic paramagnet remains one of the outstanding challenges in strongly interacting physics.

When antiferromagnetic interactions are short-ranged, frustration relies on geometry: for example, lattices containing plaquettes with an odd number of sites may frustrate N\'eel order.
This route is most pertinent in solid-state magnets, where exchange interactions are short-ranged, and has led to the discovery of a number of exciting spin liquid candidates  in layered two-dimensional Mott insulators \cite{Balents2010,Shimizu2003, Itou2008, RigolSingh, Helton}.
An alternate route to frustration is provided by longer range interactions.
An array of numerical studies have demonstrated that adding  further-neighbor couplings can destabilize classical order and lead to spin liquid phases. 
Unfortunately, liquid phases are often found only for a narrow range of further neighbor couplings comparable to the nearest neighbor exchange, making it challenging to identify relevant physical systems.

The recent emergence of polar-molecular gases opens a new route toward long-range interactions  \cite{ni2008high, chotia2012long,Deiglmayr08,Park15}:  in contrast to both their atomic cousins and conventional quantum materials, polar molecules exhibit strong, dipolar interactions \cite{yan2013observation,hazzard2014many,baranov2012condensed}. 
However, these interactions are neither isotropic nor obviously frustrated, leading to many proposals which `engineer' frustrated phases via the use of  multiple molecular states, microwave dressing fields, and spatially varying optical potentials \cite{lewenstein2006atomic,micheli2006toolbox,gorshkov2011tunable,gorshkov2011quantum,yao2013realizing,manmana2013topological}. 

\begin{figure}[h]
\begin{center}
\includegraphics[width=3.4in]{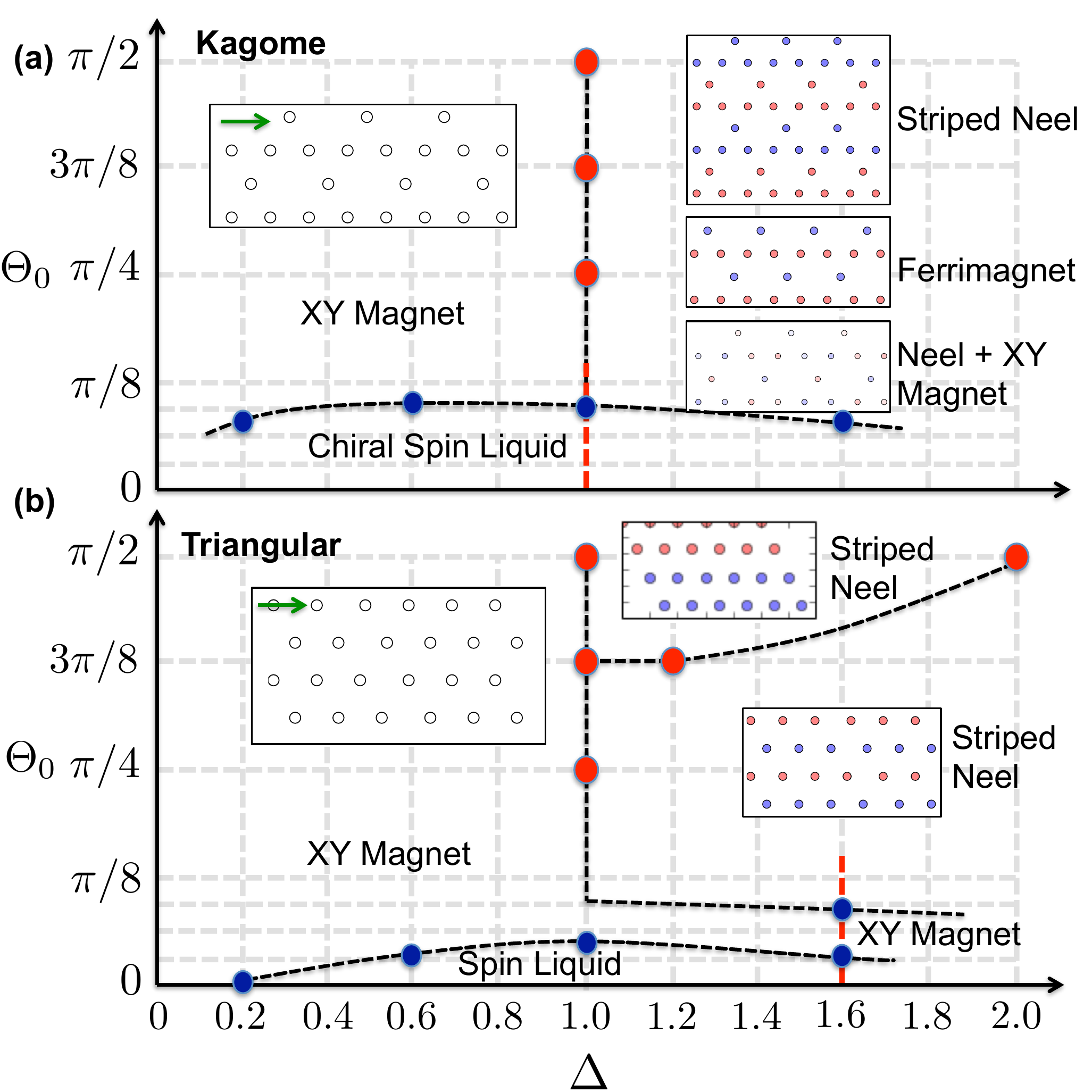}
\end{center}
\caption{ {\bf a)} Phase diagram of the dipolar Heisenberg model on the a) kagome lattice (YC8 geometry truncated at $J_8$) and the {\bf b)} triangular lattice (YC6 geometry truncated at $J_5$) as a function of the XXZ anisotropy $\Delta$ (which is controlled by the magnitude of the applied electric field [see Fig.~2]) and the polar tilt, $\Theta_0$,  of the applied electric field (the azimuthal angle is given by the green arrow). Near $\Theta_0=0$, where the model is fully frustrated, we observe quantum spin liquid ground states on both geometries. Ordered phases for $\Theta_0>0$ are shown with their corresponding magnetization density. }
\label{fig:phase}
\end{figure}

Furthermore, although long-ranged, the dipolar couplings are not easily  fine-tuned; rather,  scale invariance  dictates  that the simplest effective Hamiltonian one could hope for is a `dipolar Heisenberg antiferromagnet:'
\begin{equation}
H =  \sum_{i, j } \frac{\vec{S_i} \cdot \vec{S_j}}{|R_{ij}|^3}.
\label{eq:Heis}
\end{equation} 
Two fundamental questions arise: is $H$ naturally realized, and what is its ground state?

Here, we answer both of these questions.
First, we consider synthetic quantum magnets constructed from an array of lattice-trapped, polar molecules interacting via dipole-dipole interactions. We demonstrate that this system easily realizes the dipolar Heisenberg antiferromagnet, requiring only a judicious choice of two molecular rotational states (to represent a pseudo-spin) and  a constant electric field \cite{gorshkov2011quantum}.
The simplicity of our proposal stems from using rotational states with no angular momentum about the electric field axis. This contrasts with previous works where non-zero matrix elements appear for the transverse electric dipole operator, unavoidably generating ferromagnetic spin-spin interactions  because of the inherent anisotropy of the dipolar interaction \cite{yao2013realizing,manmana2013topological}.

Second, motivated by this physical construction, we perform a large-scale density matrix renormalization group \cite{White:1992, McCulloch:2008} and exact diagonalization study of  the dipolar Heisenberg model and find evidence for  quantum spin liquid ground states on  both  triangular and kagome lattices (Fig.~\ref{fig:phase}). 
Because of the long-range interactions and the need for time-reversal breaking complex wavefunctions, our model is one to two orders of magnitude more challenging to simulate numerically than earlier nearest-neighbor models. 
The further-neighbor dipolar couplings play a crucial role, leading to a \emph{different} phase of matter for both lattice geometries when compared to their nearest-neighbor counterparts realized in Mott insulating materials.
We compute the  phase diagram of the dipolar Heisenberg model as a function of  experimental parameters (the electric field strength and tilt) for any ultracold polar molecule.

\begin{figure}
\begin{center}
\includegraphics[width=2.7in]{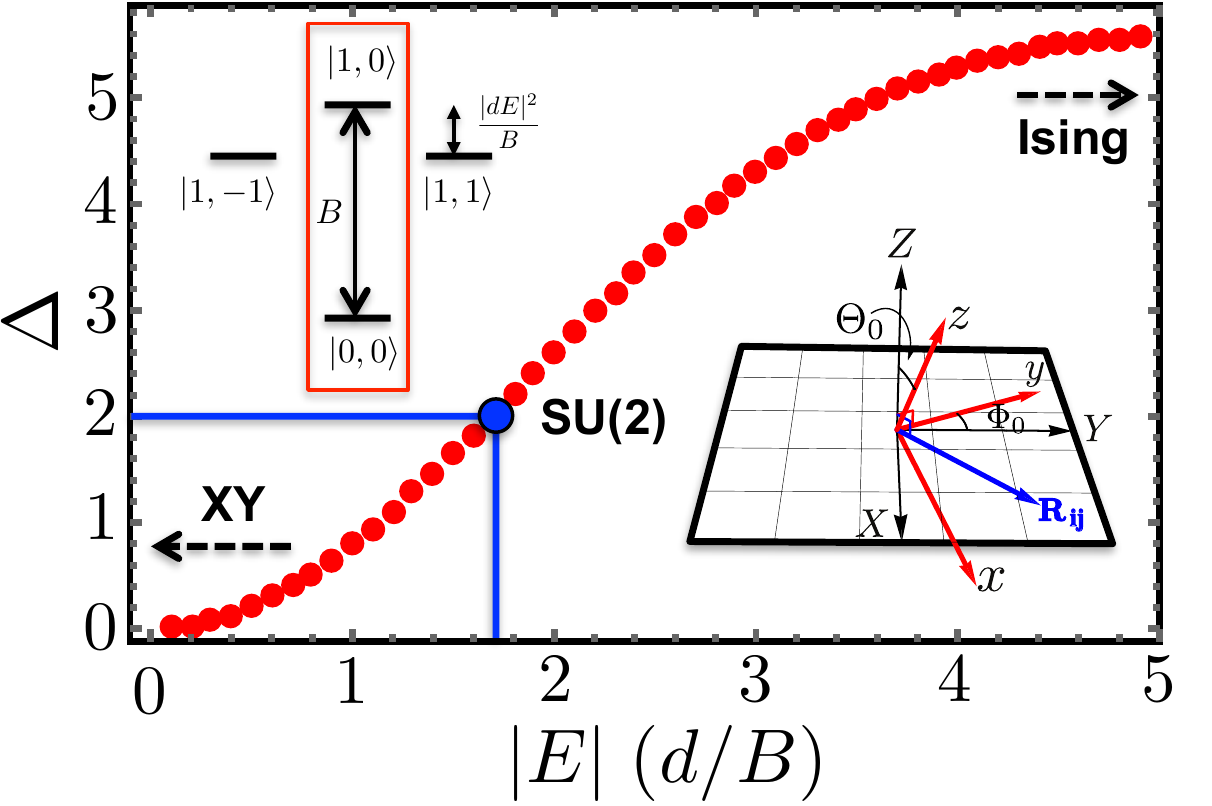}
\end{center}
\caption{The XXZ anisotropy $\Delta$ is controlled by the electric field strength, $E$, measured relative 
to the rotational splitting divided by the  dipole moment, $B/d$. Top left inset - the rotational states used as the two-level pseudo-spin. Bottom right inset - molecules reside in the XY plane and the electric field is oriented along $\hat{z}$.}
\label{fig:scheme}
\end{figure}

\emph{Realization}---We consider a two-dimensional array of  polar molecules trapped in an optical lattice.
The lattice freezes the translational motion, leaving each molecule to behave as a simple dipolar rigid rotor \cite{micheli2006toolbox,gorshkov2011tunable,gorshkov2011quantum,yao2013realizing,manmana2013topological}.
The Hamiltonian governing these molecular rotations is  $H_{m} = B \mathbf{J}^2 + \vec{E}\cdot\mathbf{d}$, where $B$ is the rotational constant, $ \mathbf{J}$ is the angular momentum operator,  $\vec{E}$ is the external electric field, and $\mathbf{d}$ is the dipole operator. 
For $|E| = 0$, each molecule  has eigenstates indexed by $\ket{J, M}$, where $M$ is the $z$-component of angular momentum.
An applied electric field, $\vec{E}  = E \hat{z}$, weakly aligns the molecules along the field direction, mixing states with identical $M$. Each $\ket{J,M}$ evolves adiabatically with $E$, picking up a dipole moment and splitting the degeneracy within each $J$ manifold at order $(dE)^2 / B$ (inset Fig.~\ref{fig:scheme}).

The molecules interact with one another via the electric dipole-dipole interaction, 
\begin{equation}
H_{\textrm{dd}} = \frac{g}{2} \sum_{i\neq j}  \frac{1}{R_{ij}^3}   \left [  {\bf d}_i  \cdot {\bf d}_j - 3({\bf d}_i \cdot {\bf \hat{R}}_{ij})({\bf d}_j \cdot {\bf \hat{R}}_{ij}) \right ],
\end{equation} 
where  $g = 1/(4\pi\epsilon_0)$ and ${\bf R}_{ij}$ is the displacement between  molecules $i$ and $j$. 
Referring to Fig.~\ref{fig:scheme}, we select the doublet  $\ket{\downarrow} = |0,0\rangle$ and  $\ket{\uparrow} = |1,0\rangle$, which are energetically resolved from all other rotational states, to play the role of a ``spin'' \cite{gorshkov2011quantum}.
We let $S^{\mu}$ denote the usual spin operators in this subspace, but note, that unlike $S=1/2$ moments, this doublet a priori lacks SO(3) symmetry.
To derive the effective Hamiltonian, we project $H_{\textrm{dd}}$ onto the two-level subspace  and drop $S^z$ non-conserving terms as they are strongly off-resonant. 
This projection is physically justified by the separation of energy scales between the dipolar interaction and the  rotational level-splittings: $g d^2 / R^3 \ll B, (d \,E)^2 / B $.

When the electric field is aligned perpendicular to the lattice plane ($\Theta_0 =0$, inset Fig.~\ref{fig:scheme}), we find \cite{gorshkov2011quantum}
\begin{equation}
H_{\textrm{eff}} = g  \sum_{i, j }  \frac{1}{R_{ij}^3}   \left[  2d_{00}^2(S^x_i S^x_j + S^y_j S^y_i) + (\mu_0 - d_0)^2 S^z_i S^z_j \right]
\label{eq:XXZ}
\end{equation} 
where $d_{00} = \langle 1,0|d_z |0,0\rangle$ is the transition dipole moment and $d_{0} = \langle 0,0|d_z |0,0\rangle$,   $\mu_0=\langle 1, 0 |d_z |1, 0 \rangle$ are the electric field induced ``permanent'' dipole moments. 
The sign of the couplings shows that the interaction is anti-ferromagnetic along \emph{all} spin-axes.

As depicted in Fig.~\ref{fig:scheme}, the ratio $\Delta = \frac{(\mu_0 - d_0)^2}{d_{00}^2}$, between the Ising and XY  interactions [Eqn.~3] is  controlled by the magnitude of the applied electric field. SO(3) symmetry emerges for $|dE| \approx 1.7B$, at which point the effective Hamiltonian is precisely the dipolar Heisenberg model.
We note that $H_{\textrm{eff}}$  is in stark contrast to the typical spin models analyzed for polar molecules.
In particular, previous works have generally considered rotational states that lead to ferromagnetic interactions favoring easy-plane (XY) magnetism; frustrated phases arise only upon fine-tuning via microwave and optical dressing \cite{gorshkov2011tunable,gorshkov2011quantum,yao2013realizing,manmana2013topological,micheli2006toolbox}.

\emph{Ground State of the Dipolar Heisenberg Antiferromagnet}---While the antiferromagnetic dipolar Heisenberg interaction is frustrated on any lattice, geometries with triangular motifs typically enhance this frustration as it is impossible for all neighboring spins to anti-align.
Here, we consider kagome and triangular lattices, both of which have been realized in optical lattices \cite{jo2012ultracold,becker2010ultracold,struck2011quantum}.

\begin{figure*}[t]
\begin{center}
\includegraphics[width=6.4in]{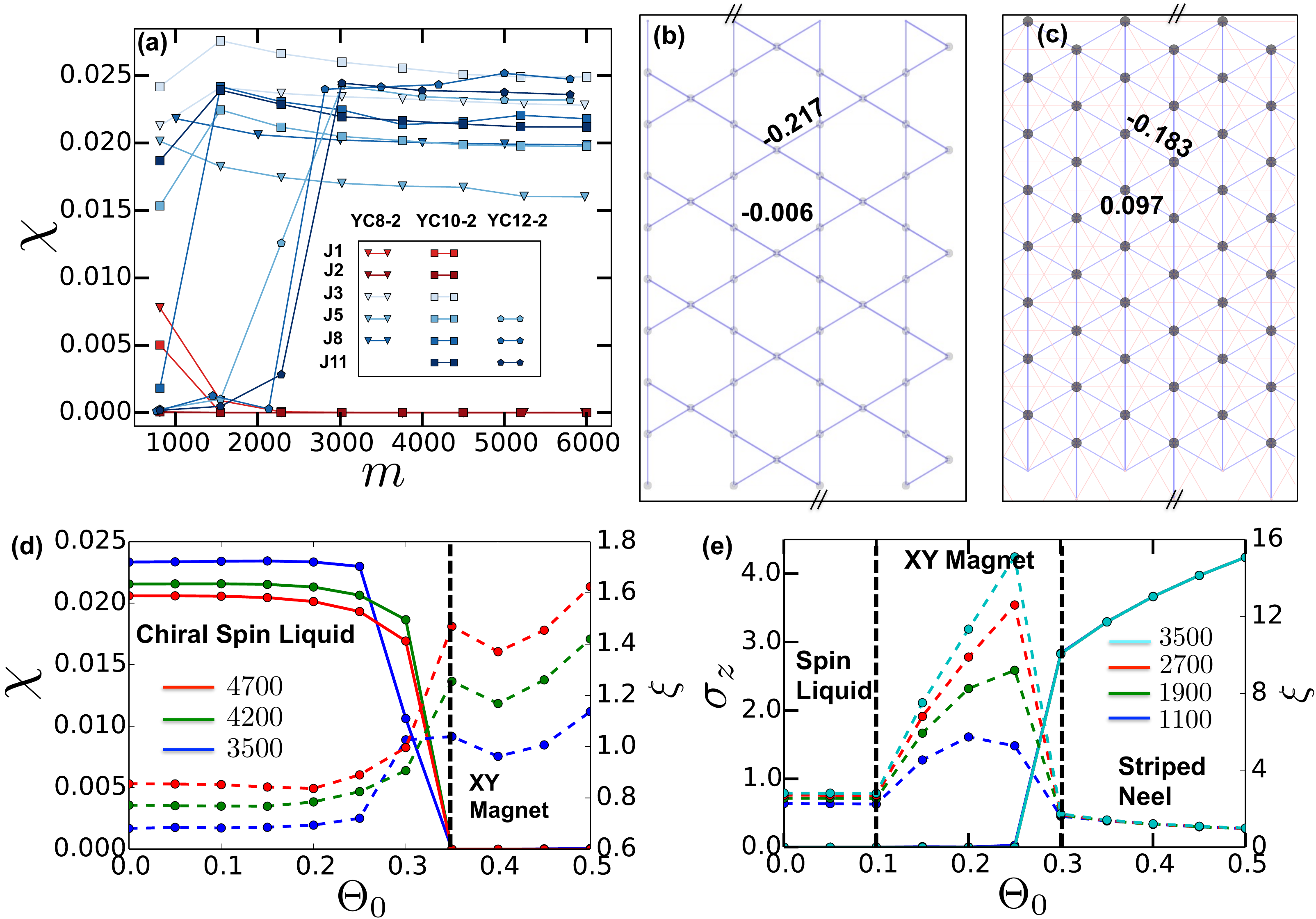}
\end{center}
\caption{\textbf{a}) Triple extrapolation of the chiral order parameter $\chi$ as a function of the DMRG bond dimension ($m$), the cylinder circumference and the range of the dipolar interaction.  For YC8-2 and YC10-2, all numerics have converged to a truncation error $<10^{-5}$ while for YC12-2, we observe a truncation error $\sim 4.5 \times 10^{-5}$ at bond dimension $m=5800$. \textbf{b},\textbf{c}) The NN and NNN $\langle S_i \cdot S_j \rangle$ correlations of the kagome (YC10-2) / triangle (YC8) spin liquid respectively. The magnitude of the correlation function for each bond is shown and is directly proportional to the linewidth of the bond  (see supplementary information for further detail). \textbf{d}) Phase transition out of the chiral spin liquid (holding $\Delta = 1.0$ fixed and varying $\Theta_0$) as characterized by the vanishing of $\chi$ and the diverging correlation length  $\xi$ (with $m$). \textbf{e}) Phase transition out of the triangular spin liquid (holding $\Delta = 1.6$ fixed and varying $\Theta_0$) as characterized  by $\sigma_z$, the variance of $S^z$ across the unit cell, and the correlation length $\xi$. }
\label{fig:results}
\end{figure*}

The ground state of  the dipolar Heisenberg antiferromagnet is unknown for either lattice.
Even for short-range interactions, the  phase diagram in these geometries has been an open question for more than two decades, due to delicate energetic competition between many competing phases.
Recently,  progress has been madeusing the density matrix renormalization group (DMRG) \cite{Yan2011, Depenbrock2012, Jiang2012,Gong15,HeChenYan2014, Gong2014, He2014, Zhu15, Hu2015}.
As DMRG is a 1D method, it requires mapping the 2D lattice to a quasi-1D geometry; here, we study both finite-length and infinitely long  cylinders of circumference $L$.  
The dipolar interaction introduces an additional difficulty, as its range  must be truncated for a consistent definition on the cylinder.
Thus, our numerics require a triple extrapolation in $L$, the interaction range, and the accuracy of the DMRG as quantified by the `bond dimension' $m$.

 \begin{figure*}[t]
\begin{center}
\includegraphics[width=7.0in]{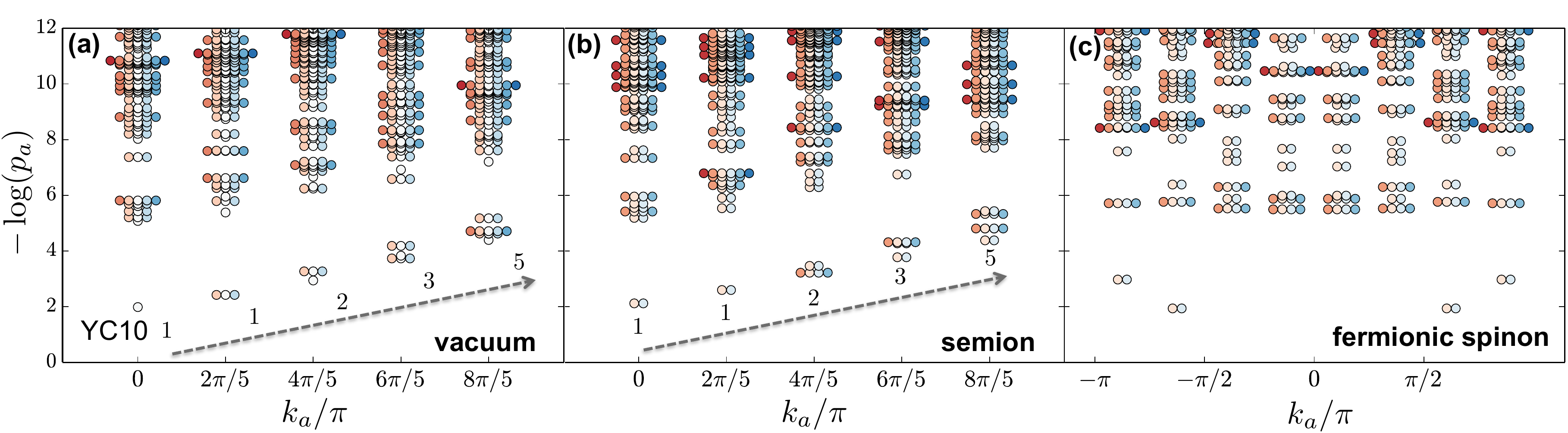}
\end{center}
\caption{The entanglement spectrum $\{p_a\}$ as a function of $k_a$, the momentum of the Schmidt state around the cylinder. Data points are colored and displaced slightly according to their $S^z$ quantum numbers. 
 \textbf{a}) The \emph{kagome} YC10 model truncated at $J_8$. The momentum-resolved entanglement spectrum is consistent with the vacuum sector of a chiral spin liquid and exhibits the characteristic counting $\{1, 1, 2, 3, \cdots \}$ predicted by the WZW edge theory.  \textbf{b}) Shifting the cut of the YC10 cylinder by a single column gives the semion sector of the spin liquid, with half-integral representations of SO(3).   \textbf{c}) The \emph{triangular} YC8 model truncated at $J_5$. The spectrum is consistent with the fermionic spinon topological sector of a $\mathbb{Z}_2$ spin liquid.
 }
\label{fig:espectrum}
\end{figure*}

Detecting and characterizing a quantum spin liquid phase follows a decision tree. 
By definition, ``liquid'' refers to the absence of  spontaneous symmetry breaking, specifically of spin-rotations and translation invariance. 
Any liquid phase with half integer spin in the unit cell \emph{must} be exotic: the Hastings-Oshikawa-Lieb-Schultz-Mattis theorem requires that the phase  be either an exotic gapless spin liquid  or a gapped  spin liquid with fractionalized excitations \cite{Oshikawa2000, Hastings2005}.
In the gapless case, the ground state has a diverging correlation length as the circumference of the cylinder is increased.
In the gapped case, the ground state will have exponentially decaying correlations, protected ground state degeneracy, and certain characteristic signatures in its  entanglement spectrum \cite{KitaevPreskill, zaletel2015measuring}.

There exists a zoo of gapped spin liquids distinguished by the braiding and statistics of their fractional excitations.
The two simplest cases are the time-reversal symmetric (TRS) $\mathbb{Z}_2$ spin liquid and the time-reversal breaking chiral spin liquid (CSL) \cite{Kalmeyer87, Wen89};  the spontaneous breaking of time-reversal is detected by using a chiral order parameter $\chi =  \langle S_i \cdot S_j \times S_k \rangle / 3$, where $i, j, k$ are the three sites of a triangle. 

Let us now turn to the numerics.
We refer to the cylinder geometries using the notation of [\onlinecite{Yan2011}]; YCn is a cylinder of circumference $n$ lattice spacings periodized along a Bravais vector (Fig.~\ref{fig:results}b,c). 
For both lattices, we define $J_n$ to be the coupling between $n$\textsuperscript{th} nearest neighbor sites, ordered by their distance  in real space, $R_n$. We will begin by characterizing the ground state of each lattice at the dipolar Heisenberg point and will subsequently map out the full phase diagram of the molecular proposal.

\emph{Kagome Model}---
Extensive theoretical and numerical studies  of the $J_1-J_2-J_3$ kagome model reveal a rich phase diagram, consisting of a honeycomb valence bond solid, a $\mathbb{Z}_2$ spin liquid, a chiral spin liquid, and a multitude of ordered N\'eel states \cite{singh2007ground, lauchli2011ground, Yan2011, Depenbrock2012, Jiang2012,Gong15,HeChenYan2014, bauer2014chiral, Gong2014, He2014}.
In contrast to these previous studies, the long-range dipolar couplings cannot be tuned. For the kagome lattice it is necessary to distinguish between two couplings of length $R_3=2a$: $J_3$ (across hexagons) and $J'_3$ (along bow-ties). Motivated by exchange interactions in Mott insulating materials, previous numerics have always considered $J'_3=0$. In the dipolar Heisenberg model, all couplings at a given distance are equally important and a finite $J_3'$ in fact  stabilizes the CSL phase (see supplemental information for details). This is highlighted by the fact that keeping only the $J_2$ or $J_3$ part of the dipolar interaction results in the magnetically ordered $\mathbf{q} = (0, 0)$ phase  \cite{Gong15,HeChenYan2014, Gong2014, He2014}; only upon restoring the dipolar tail of the interaction does the system transition into the CSL. 

Let us now turn to the  diagnostics of liquidity. We study cylinders of circumference $L = 8, 10, 12$ with dipolar cutoffs ranging from $J_3$ to $J_{11}$. In addition to the YCn geometry, we also consider  the so-called `YCn-2'  geometry in which cylinders  are rolled up with a `twist' that identifies sites that differ by Bravais vector $n \vec{a}_1 + \vec{a}_2$. This convenient choice of boundary condition reduces the computational cost by decreasing the effective iDMRG unit cell, enabling better convergence for certain diagnostics. Crucially, neither the spin liquids nor the $\mathbf{q} = (0, 0)$ phase are frustrated by this boundary condition; more generally, for liquid phases, the resulting physics should be unaffected once the cylinder circumference is larger than the correlation length.

We find $\langle S^{\mu} \rangle = 0$, as required by the Mermin-Wagner theorem in our quasi-1D geometry (note that for $\Delta > 1$, spontaneous N\'eel order is allowed, but  not observed).
A tendency towards N\'eel order should appear as algebraic correlations beyond the dipolar cutoff; instead we find a short correlation length $\xi \lesssim 0.9a$ (as calculated from the DMRG transfer matrix), consistent with a gapped paramagnet. 
The absence of local magnetization and long-range correlations indicates that spin rotation symmetry is preserved.

To check that translational symmetry is also preserved (i.e.~to rule out valence-bond order), we verify that the bond correlations are translation invariant (Fig.~\ref{fig:results}b) and also calculate the overlap of the ground state, $\ket{\Psi}$, with a translated version of itself, $\bra{\Psi}  \hat{T}^y \ket{\Psi} = (1 - \epsilon)^{V} $. This overlap scales with the volume of the system, $V$, with error $\epsilon < 0.004$. 
The above are quoted for a YC10 geometry with couplings up to $J_8$, but similar results are found when truncating to $J_3$ or extending to $J_{11}$, as well as on the smaller YC8 geometry and the larger YC12-2 geometry (see supplementary information for  details).

A key indication of the CSL phase is the spontaneous breaking of time-reversal symmetry.
To this end, the chiral order parameter $|\chi |$ is shown in Fig.~\ref{fig:results}a as a function of the size of the cylinder, the truncation cutoff, and the DMRG accuracy;
$|\chi |$ increases weakly with  cylinder circumference, converges  with bond dimension, and saturates for large dipolar cutoff. 

In addition to spontaneous TRS breaking, the most spectacular signature of a CSL is a chiral edge state.
Quantum entanglement provides a  way to probe these edge states given only  the ground state.
The reduced density matrix $\rho_L$ for half of the cylinder can be viewed as a thermal density matrix of a semi-infinite cylinder, introducing a single `edge'.
The spectrum $p_a$ of $\rho_L$ (the `entanglement spectrum') is known to mimic the energy spectrum of the physical edge; plotting this spectrum versus the momentum around the edge, $k_a$, should  reveal a chiral dispersion relation \cite{KitaevPreskill, LiHaldane,QiKatsuraLudwig}.
As shown in Fig.~\ref{fig:espectrum}a,b the momentum-resolved entanglement spectrum of a YC10 cylinder indeed displays characteristic  level counting $\{1, 1, 2, 3, 5 \cdots\}$ organized into SO(3) multiplets consistent with the SU(2)$_1$ Wess-Zumino-Witten edge theory \cite{wess1971consequences}.

\emph{Triangular Model}---Truncating the dipolar Heisenberg model at short range leads to N\'eel order: for $J_1$ only, a 120$^\text{o}$ degree N\'eel phase  \cite{Jolicoeur90}, and for $J_1, J_2$, a  two-sublattice collinear N\'eel phase \cite{Zhu15, Hu2015}. 
However, adding in the dipolar $J_3$ coupling directly penalizes the order of the collinear state and appears to drive the system into a liquid; this is evidenced by a drastic change in the $\langle S_i \cdot S_j \rangle$ correlation function as the long-range tail of the  interaction is restored (see supplementary information).
With couplings through $J_5$, the YC8 ground state has an XY correlation length of $\xi \lesssim 1.4a$ and is translationally symmetric with $\epsilon < 4\times 10^{-5}$.
Similar results are found when truncating to $J_3$ or extending to $J_8$, as well as on the smaller YC6 geometry and the larger YC10 geometry.

The phenomenology of the observed spin liquid phase is equivalent to the $J_1-J_2$ spin liquid reported in \cite{Zhu15, Hu2015}.
The lowest energy state is time-reversal symmetric and has an  entanglement spectrum consistent with the fermionic spinon topological sector of a $\mathbb{Z}_2$ spin liquid; it exhibits a four-fold degeneracy and a half-integral representation of SO(3) as shown in Fig.~\ref{fig:espectrum}c \cite{zaletel2015measuring}.
While the bond correlations are translation invariant (Fig.~\ref{fig:results}c), they exhibit a noticeable striping consistent with  nematic ordering (note that this nematicity may be an artifact of the cylindrical geometry  which breaks  $C_6$ symmetry) \cite{Zhu15}.

\emph{Phase Diagram}---The above results (for both triangular and kagome) were presented for the SO(3) symmetric Heisenberg anti-ferromagnet ($\Delta = 1$) at $|dE| \approx 1.7B$.
For both stronger ($\Delta = 1.6$) and weaker  ($\Delta = 0.6$) electric fields, the SO(3) model is broken down to a U(1) XXZ model, but our numerics find the spin liquid phases are  completely consistent with those observed at the SO(3) point \cite{He2014}.
Note that the Hastings-Oshikawa-Lieb-Schultz-Mattis theorem requires only U(1) invariance about the $z$-axis and zero net magnetization.

As one tilts the electric field into the lattice plane, the spin liquids we observe begin to  compete with magnetically ordered phases. The tilt generates angular dependence in the effective Hamiltonian,
\begin{align}
H_{\textrm{eff}} =  g \sum_{i, j } \frac{1}{R_{ij}^3} [ 1 - 3 \cos^2(\Phi - \Phi_0) \sin^2 \Theta_0]  \times \nonumber \\ [2d_{00}^2(S^x_i S^x_j + S^y_j S^y_i) + (\mu_0- d_0)^2 S^z_i S^z_j] 
\end{align}
where $\Phi$, $\Phi_0$ are the polar angles of  $\vec{R}_{ij}$ and the electric field orientation, respectively (inset of Fig.~\ref{fig:scheme}).
For nonzero $\Theta_0$, full frustration is lost as dipoles begin to point head-to-tail along the field direction, thereby exhibiting  ferromagnetic interactions. 
For large $\Theta_0$, a variety of ordered phases appear as shown in Fig.~\ref{fig:phase}a,b (for full details, see supplementary information). Here, we restrict our interest to the phase boundaries  of the spin liquid states.

In Fig.~\ref{fig:results}d,e, we present two representative vertical cuts: 1) out of the kagome CSL at $\Delta =1.0$ and 2) out of the triangular spin liquid at $\Delta = 1.6$. In the kagome case, we identify the transition out of the chiral spin liquid  via the vanishing of the chiral order parameter (Fig.~\ref{fig:results}d). % and the diverging of the correlation length as a function of $m$ 
In the triangular case, we  diagnose the phase transition  by examining the correlation length and the variance of the $S^z$-magnetization (Fig.~\ref{fig:results}e). This reveals two phases, an XY magnet directly proximate to the spin liquid and the expected striped N\'eel phase for larger $\Theta_0$.
In addition to showing that the spin liquid phases persist to moderate electric field tilts, understanding the nature of the ordered phases surrounding the spin liquids may enable the preparation of these topological states \cite{barkeshli2015continuous}.

In summary, our proposal provides a new route toward studying frustrated quantum magnetism in an ultracold lattice gas. The dipolar Heisenberg antiferromagnet exhibits promising signs of spin liquid behavior on both the kagome and triangular lattices, distinct from models of nearest-neighbor exchange. Looking forward, it is important to consider the effects of lattice vacancies and dipolar relaxation as well as to identify unique signals of frustration in quench dynamics.

We gratefully acknowledge the insights of and discussions with B. Lev, A. Gorshkov, A. M. Rey, M. Lukin, C. Laumann, J. Moore, M. Zwierlein and J. Ye. This work was supported by the AFOSR MURI grant FA9550-14-1-0035 and the Miller Institute for Basic Research in Science.

\bibliography{references}

\newpage

\end{document}

\hspace{30mm}{\bf \large Supplementary Information for A Quantum Dipolar Spin Liquid}
\vspace{5mm}

\hspace{45mm}N. Y. Yao, M. P. Zaletel, D. M. Stamper-Kurn, A. Vishwanath

\vspace{15mm}

\noindent Here, we present additional numerics for both the kagome and triangular models. We also provide a description of the ordered phases that appear at large electric field tilts $\Theta_0$ and examples of phase transitions among them.

\vspace{5mm}

{\bf Kagome lattice}---We computed the ground state on geometries YC8, YC10, YC12 and YC8-2, YC10-2, YC12-2 (`shifted'). 
The topologically degenerate ground states of a spin-liquid on an un-shifted even circumference cylinder (e.g. YC8) are of two types: `odd' sectors, in which Schmidt states carry  half-integral representations of SO(3), and the `even' sectors, in which they carry integer representations.
For YC8, we find ground states in both the even and odd sectors.
For odd circumference and shifted cylinders, there is no such distinction.
Calculations were repeated for a dipolar cutoff at $J_3$, $J_5$, $J_8$ and $J_{11}$.
There is a tradeoff between the number of couplings kept and how well we can converge the DMRG in the bond dimension $m$.
In the main text, we presented data for the shifted geometries for cutoffs from $J_3$ to $J_{11}$. 
\begin{figure}[h]
\includegraphics[width=6.0in]{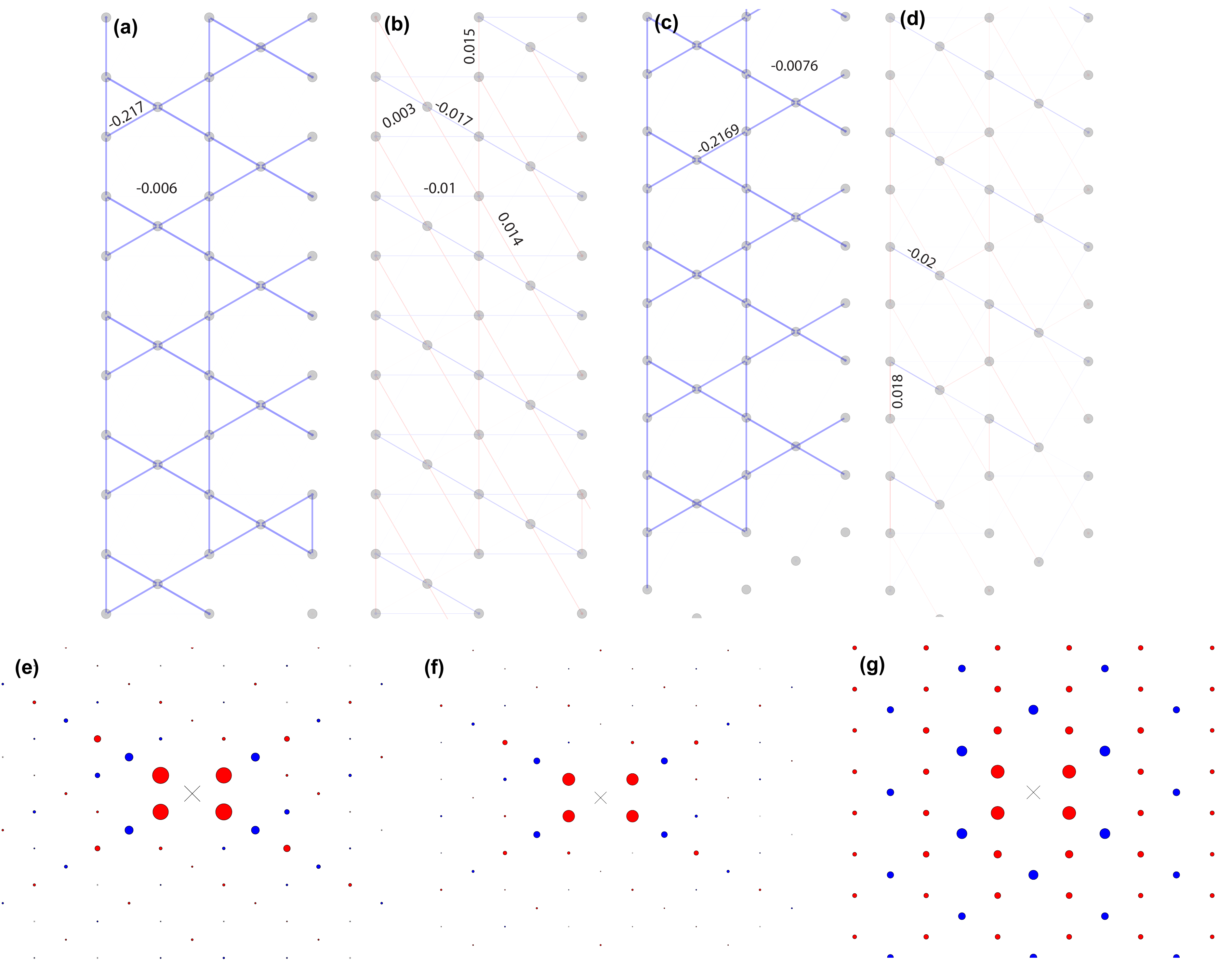}
\captionsetup{justification=raggedright,
singlelinecheck=false
}
	\caption{  \textbf{a, b}) The NN and NNN $\langle S_i \cdot S_j\rangle$ correlation function of the YC10-2 kagome $J_{1}- J_8$ model. Data is taken at $m = 6000$. In panel (a) we show $\langle S_i \cdot S_j\rangle$, with  blue negative and red positive; magnitudes of two typical bonds are labeled. In panel b), we plot the correlations after subtracting off the mean value for the bond type, revealing deviations. There is apparently a slight anisotropy, which is expected from the combination of a chiral order and the `twisted' nature of the YC10-2 cylinder.
\textbf{c-d}) same as in (a,b) for the YC10 kagome model. Unfortunately, we cannot fully converge the YC10 geometry ($m = 3600$), leading to a mottled pattern in the correlations at the level of 10\%.
Within this error results are consistent with YC10-2.
\textbf{e}) The $\langle S_0 \cdot S_i\rangle$ correlation function of the YC10-2 model; site $0$ is indicated with an $\times$. \textbf{f}) The $\langle S_0 \cdot S_i\rangle$ correlation function of the YC10  mode. \textbf{g}) For contrast, the $\langle S_0 \cdot S_i\rangle$ correlation function of the dipolar YC8 model truncated at $J_1 -J_2$,  which is known to have N\'eel order. Data is plotted on the same scale as (e), (f).
    }
	\label{fig:fig1_v1_schem}
\end{figure}
Here, since we discuss the more difficult YC8 and YC10 geometries, we focus on the $J_8$ cutoff.
We address two questions: 1) does the system have spontaneous chiral order? 2) is the state a liquid?

\paragraph{Chiral order.}
All DMRG simulations were done with complex wavefunctions in order to allow for spontaneous chiral order.
For every sample except \emph{one} - the even sector of YC8 - the state breaks time reversal with chiral order parameter $\chi \sim 0.02$. 
Data for all the shifted samples was summarized in the main text, and strongly suggests that chiral order persists in the thermodynamic limit.
The YC10 chiral order is also $\chi \sim 0.023$, consistent with the shifted result.
For the exceptional case, the YC8 model truncated at $J_8$, the energies are $E_{\textrm{YC8:E}} = -0.4296$ (with $\chi = 0$) and $E_{\textrm{YC8:O}} = -0.4308$ (with $\chi = 0.019$) at $m = 4200$. 
The YC8 odd sector has a chiral entanglement spectrum characteristic of the CSL.
To interpret the discrepancy between the odd and even cases, note the splitting of the topological degeneracy  in the TR-symmetric nearest neighbor YC8 model is $\Delta E \sim 0.0005$, with the even sector lowest in energy.
One interpretation is that the energies of the CSL and a TR symmetric SL are split by about $\Delta E \sim 0.001$ per site; for the YC8 cylinder, the splitting of the topological degeneracy is comparable to this competition, and we find one state from each phase.
For the other cylinders, there is no odd-even distinction, and we always find a CSL phase. 
It would be useful  to study the odd / even sectors of the YC12 cylinder to verify this hypothesis.
We are unable to converge YC12 for the $J_5$ and $J_8$ models, so must leave this to  future  SU(2) DMRG studies.

\paragraph{Evidence for a liquid.}
To assess if the state is a liquid, we focus here on the YC10-2 and YC10 cylinders.
While we can reasonably converge YC10-2 ($m = 6000$), we cannot fully converge YC10  ($m=4000$) leading to artifacts in the correlations.
Fig.~S1(a-d) depicts the NN and NNN valence-bond correlations.
Note that these geometries are expected to have a very slight doubling of the unit cell, which should decrease exponentially with the cylinder circumference; indeed we find the bond energies differ on the order of $\Delta E \sim 0.001$ between the two sublattices, which is not observable on the scale of the figure.
Fig.~S1(e-f) shows the $S_0 \cdot S_i$ correlation function, both for YC10 and YC10-2.
%, which can be compared with Fig. 2 of the $J_1-J_2-J_3$ study in Ref.\onlinecite{gong2014emergent}. 
For contrast, we also show the $S_0 \cdot S_i$  correlation function when the dipolar model is truncated at $J_2$ [Fig.~S1(g)], which exhibits $\mathbf{q} = (0, 0)$ order.
As expected, the correlations in the $\mathbf{q} = (0, 0)$ phase are far larger, and longer ranged, than those of the putative CSL.

Finally, to illustrate the convergence of the DMRG for YC10 and YC10-2 , we plot [Fig.~S2(a,b)] the energy, correlation length, chiral order and entanglement entropy versus the MPS bond dimension $m$. Fig.~S2(d) provides a representative example of data at the smaller YC8 geometry and the larger YC12 geometry, both of which are consistent with the numerics for YC10. 

\begin{figure}
	\includegraphics[width=0.9\linewidth]{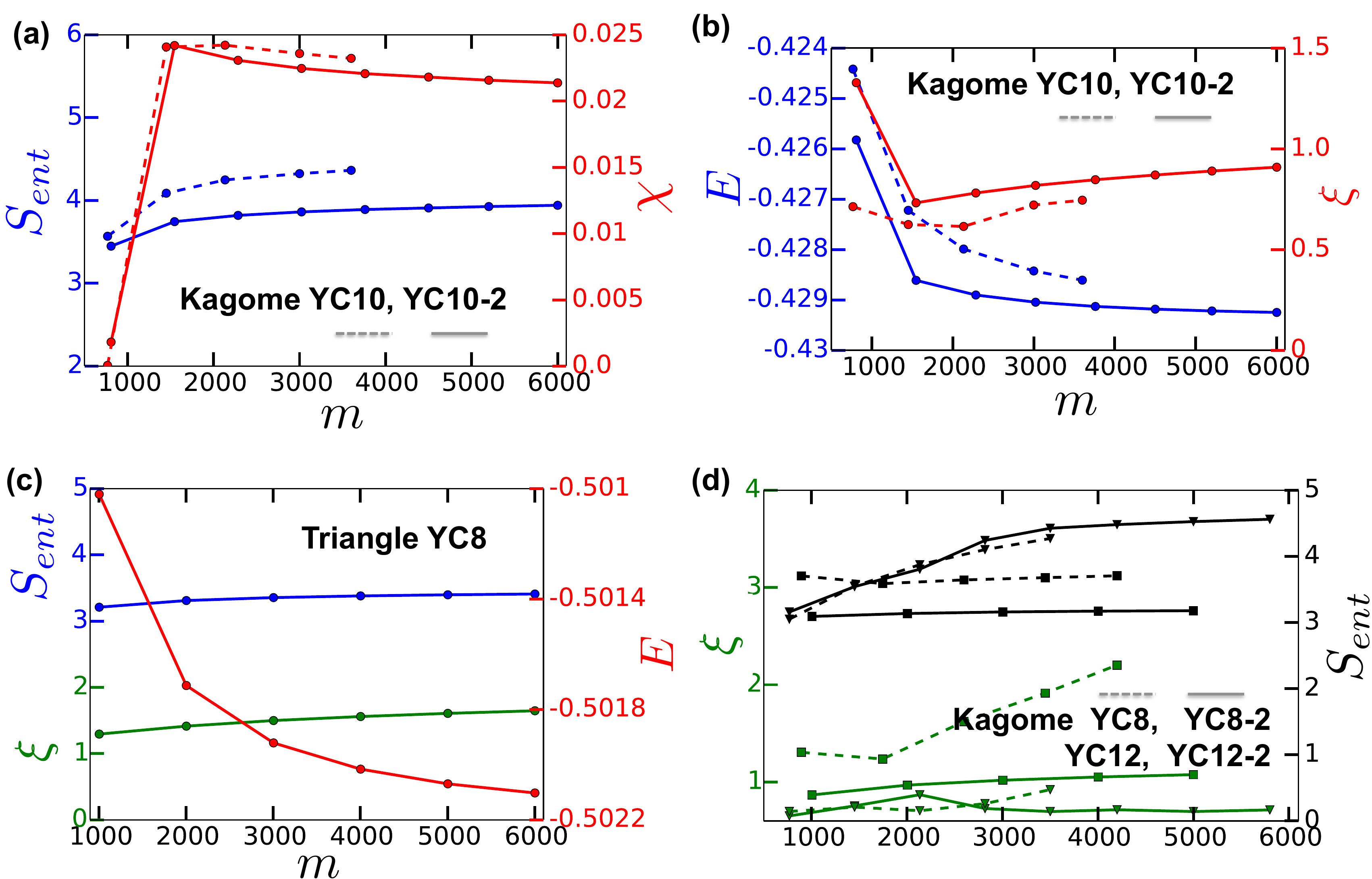}
	\captionsetup{justification=raggedright,
singlelinecheck=false
}
	\caption{(\textbf{a}-\textbf{b}) Convergence of the energy ($E$), entanglement entropy ($S_{\rm ent}$), correlation length ($\xi$) and chiral order ($\chi$) versus the MPS bond dimension $m$, for the kagome YC10 and YC10-2 models with couplings up to $J_8$. \textbf{c}) Same for the triangular YC8 model with couplings up to $J_5$ \textbf{d}) Convergence of $\xi$ and $S_{\rm ent}$ on the smaller YC8 (square markers) and larger YC12 (triangle markers) geometry for the kagome $J_8$ model. } \label{fig:pot}
\end{figure}

\vspace{5mm}

 {\bf Triangular lattice}---We first review the phenomenology found in Refs. \onlinecite{Zhu15, Hu2015}, because equivalent results (and  open questions) are found in the dipolar model. 
For YC8, the odd sector has the lowest energy, and is significantly easier to converge with bond dimensions than the even sector.

\begin{figure}[h]
\includegraphics[width=6.0in]{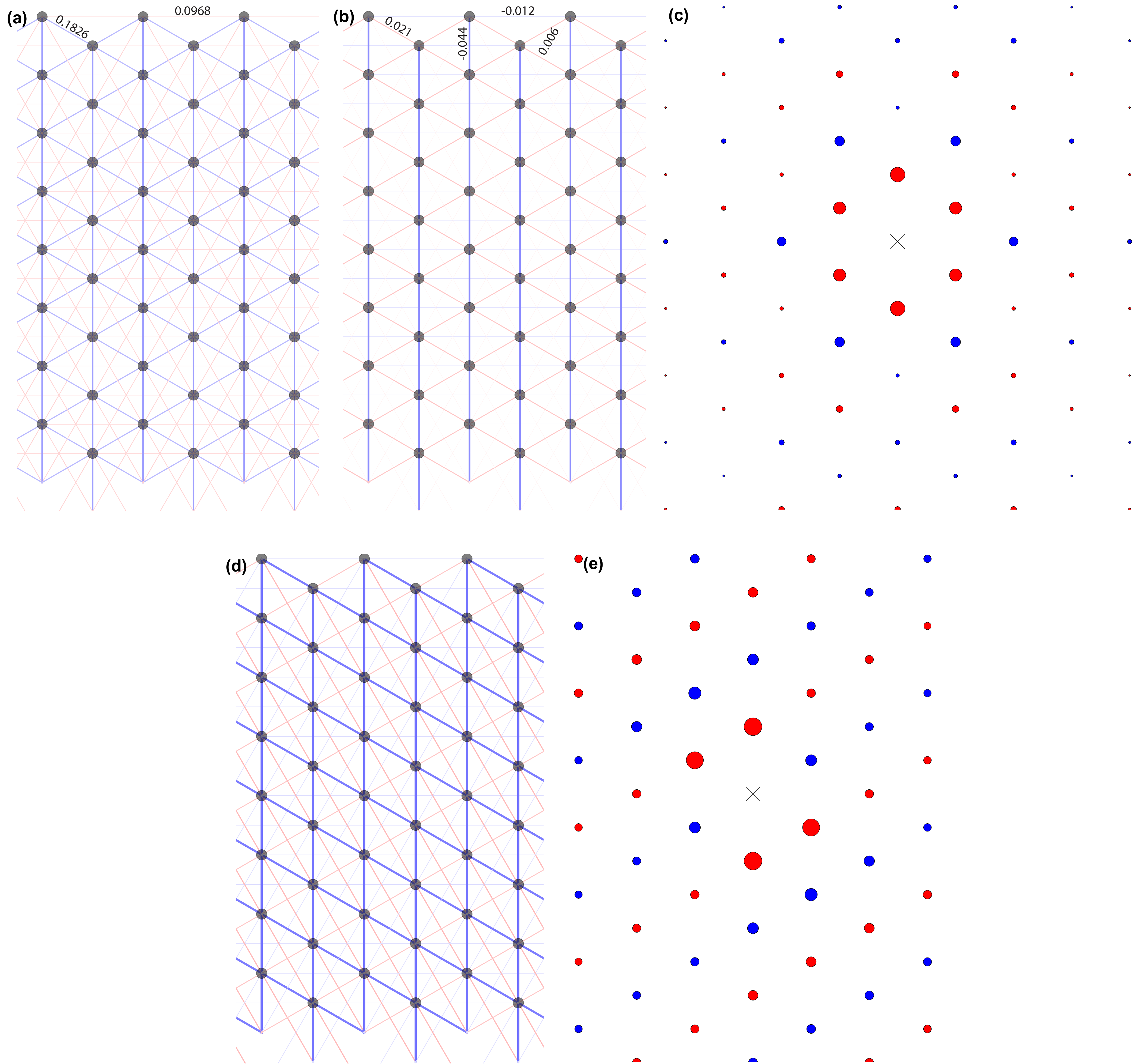}
\captionsetup{justification=raggedright,
singlelinecheck=false
}
	\caption{  \textbf{a}) The NN and NNN $\langle S_i \cdot S_j\rangle$ correlation function of the YC8 triangular $J_{1}- J_5$ model (odd sector). Data is taken at $m = 6000$. In panel $a$ we show $\langle S_i \cdot S_j\rangle$, with  blue negative and red positive; magnitudes of two typical bonds are labeled. In panel \textbf{b}), we plot the correlations after subtracting off the mean value for the bond type, revealing deviations. \textbf{c}) The $\langle S_0 \cdot S_i \rangle$ correlation function of the YC8 triangular $J_1 - J_5$ model (odd sector). Site $0$ is indicated with an $\times$. \textbf{d},\textbf{e}) Correlation functions of the triangular YC8 $J_1-J_2$ dipolar  model, which is known to be magnetically ordered. In d) we show the NN and NNN $\langle S_0 \cdot S_i \rangle$ correlation function. }
	\label{fig:fig1_v1_schem}
\end{figure}

The odd sector has 1) a short correlation length, 2) no sign of chiral ordering, and 3) a four-fold degenerate entanglement spectrum, which was argued in Ref.~\onlinecite{zaletel2015measuring} to be consistent with the `fermionic spinon' topological sector of a $\mathbb{Z}_2$ spin-liquid.
But confusion arises when considering the even sector.
In Ref.~\onlinecite{Hu2015}, it was found that the YC8 even sector has chiral-correlations out to long distances, and on YC10 it spontaneously breaks time-reversal.
This is  problematic if both sectors are to be interpreted as the degenerate ground states of a single spin-liquid.
One possibility is that there are two distinct spin liquids, one which preserves TR and one which does does not.
On these cylinders the splitting of their topological degeneracy may be comparable  to the splitting between the two spin liquids, and hence the DMRG finds one state from each.
Larger system sizes will be required to resolve this issue, which are currently beyond the reach of DMRG.

In light of this situation, our goal is simply to convince the reader that the lowest energy state of the truncated dipolar interaction on a YC8 cylinder is essentially equivalent to the fermionic spinon state observed in previous studies.
In our U(1) DMRG we are unable to study the even sector beyond bond dimensions $m > 2400$, as the simulation tunnels into the odd sector.
The entanglement spectrum of the odd sector is shown in Fig.~4c of the main text; it can be compared with Fig. 5d) of Ref.~\onlinecite{Hu2015}, which has the same 4-fold degeneracy.

In Fig.~S3(a-b) we plot the nearest and next-nearest neighbor correlations $\langle S_i \cdot S_j \rangle$, both in absolute value and relative to the mean value of the bond-type.
Most importantly, translation symmetry is preserved, which rules out valence bond crystal order at this circumference.
There is a noticeable stripe pattern; the stripes preserve translation, but on the plane would imply nematic order. 
Nematic order does not invalidate the Hastings-Lieb-Schultz-Mattis theorem.
The cylinder geometry breaks the $2 \pi / 6$ rotational symmetry, so the nematic order could be a finite size artifact.
The same stripe was found in Refs.~\cite{Zhu15, Hu2015}. In Fig.~S3(c) we plot the set of $\langle S_0 \cdot S_i\rangle$ correlations with site `$0$'.
We compare both types of correlations with the $J_1 - J_2$ only model, Fig.~S3(d-e), which is expected to be a magnetically ordered N\'eel state.
By contrast, the dipolar case shows no sign of N\'eel order.
Note that in the presence of power law interactions, the true ground state of the plane will generically have power law correlations, even if it is gapped, with a power law related to the fall-off of the Hamiltonian.
Finally, to demonstrate convergence of the DMRG for this geometry, in Fig.~\ref{fig:pot} we plot the energy, correlation length, and entanglement entropy versus the MPS bond dimension $m$.

\begin{figure*}
	\includegraphics[width=127mm]{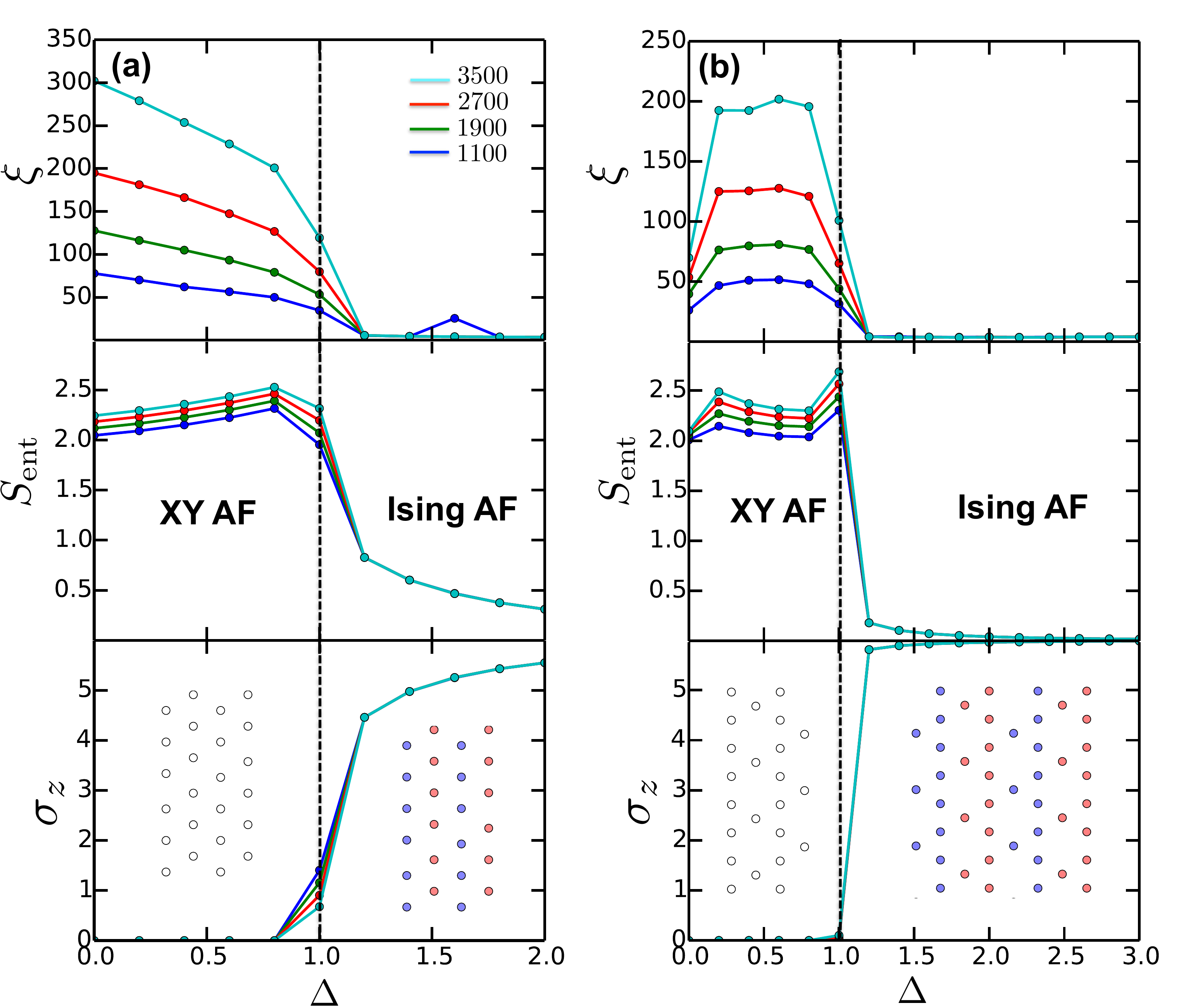}
	\captionsetup{justification=raggedright,
singlelinecheck=false
}
	\caption{a) Transition (at $\Theta_0=\pi/4$) between the XY antiferromagnet and the Ising antiferromagnet on the YC6 triangular geometry with couplings up to $J_5$. b)  Transition (at $\Theta_0=\pi/2$) between the XY antiferromagnet and the Ising antiferromagnet on the YC8 kagome geometry with couplings up to $J_8$. In both cases,  the phase transition is diagnosed by the diverging of $\xi$ in the XY phase and the onset of N\'eel order in the Ising phase.  } \label{fig:transition}
\end{figure*}

\vspace{5mm}

\noindent {\bf Order states at $\Theta_0>0$ and Associated Phase Transitions}---The nature of the ordered phases away from $\Theta_0=0$ and their phase boundaries are likely to be sensitive to the tails of the dipolar interaction. In this study, we restrict to dipolar interactions cut-off at $J_8$ (kagome) and $J_5$ (triangular) [see Fig.~1 in the maintext]. 

The nature of the ordered phases depends strongly on the XXZ anisotropy $\Delta$.
Consider first the triangular case with $\Delta > 1$. 
As large  tilts, ferromagnetism begins to develop along the field direction yielding two different striped $Z$-ordered Neel phases (maintext, Fig.~1b). 
Because of the Ising anisotropy $\Delta > 1$, this state does not break any continuous symmetries, resulting in a gapped state with low entanglement.
In contrast, for $\Delta < 1$, similar striped ordering occurs but in the XY-plane, which breaks a continuous symmetry and leads to a Goldstone boson. In the cylinder geometry studied here, the ordering is only algebraic (Mermin-Wagner theorem), consistent with the diverging correlation length found in our simulations. This distinction is exemplified by observing the phase transition across the $\Delta=1$ [Fig.~S4(a)].

The kagome model also hosts a similar array of ordered phases: for $\Delta < 1$ we find algebraically-ordered XY magnetism, while for $\Delta > 1$ we observe a number of different Neel states. A similar phase transition between these ordered states is shown in Fig.~S4(b).

\vspace{5mm}

\noindent {\bf XY Antiferromagnetism for Molecular Rotational State $|1,1\rangle$}---Working with the pseudo-spin defined by $\left |\downarrow \right \rangle = |0,0\rangle$ and $\left |\uparrow \right \rangle = |1,1\rangle$, we will demonstrate that the effective Hamiltonian naturally favors XY Neel order rather than spin liquidity.  Let us assume that the state  $\left |\uparrow \right \rangle = |1,-1\rangle$ is energetically split away, so that the only non-zero matrix elements arise from $T^2_0(\mathbf{d}^{(i)}, \mathbf{d}^{(j)})$. It is natural to  define a hardcore bosonic operator $a^\dagger_ i = \left | \uparrow \rangle \langle \downarrow \right |_i$. 

The only non-zero hopping matrix element for the rotational states is:
\ba
\bra{ \uparrow_i \downarrow_j } T^2_0 \ket{ \downarrow_i \uparrow_j} &=& - \frac{d_{01}^2 }{\sqrt{6}},
\ea
where $d_{01} = \langle 1,\pm1|d_{\pm} |0,0\rangle$ is the transition dipole moment. This yields a hardcore bosonic hopping term, $- \frac{d_{01}^2 }{\sqrt{6}}(a^\dagger_i a_j + a^\dagger_j a_i)$. The coefficient of the interaction term $n_i n_j$ depends on the induced permanent dipole moment of the rotational states, 
\ba
\bra{ \downarrow_i \downarrow_j } d_z d_z + \frac{1}{2} (d_+ d_- + d_- d_+) \ket{ \downarrow_i \downarrow_j} &=& d_0^2, \\
\bra{ \uparrow_i \downarrow_j } d_z d_z + \frac{1}{2} (d_+ d_- + d_- d_+) \ket{ \uparrow_i \downarrow_j} &=& d_{1}d_0, \\
\bra{ \downarrow_i \uparrow_j } d_z d_z + \frac{1}{2} (d_+ d_- + d_- d_+) \ket{ \downarrow_i \uparrow_j} &=& d_0d_{1}, \\
\bra{ \uparrow_i \uparrow_j } d_z d_z + \frac{1}{2} (d_+ d_- + d_- d_+) \ket{ \uparrow_i \uparrow_j} &=& d_{1}^2,
\ea
where $d_{0} = \langle 0,0|d_z |0,0\rangle$ and $d_1=\langle 1,\pm1|d_z |1,\pm1\rangle$ are the (electric field) induced dipole moments \cite{yao2013realizing}.
The effective density-density interaction strength is given by, $V_{ij}= \bra{ \uparrow_i\uparrow_j } H_{dd} \ket{ \uparrow_i\uparrow_j} + \bra{ \downarrow_i\downarrow_j } H_{dd}\ket{ \downarrow_i\downarrow_j} -  \bra{ \uparrow_i\downarrow_j } H_{dd} \ket{ \uparrow_i\downarrow_j} - \bra{ \downarrow_i\uparrow_j } H_{dd}\ket{ \downarrow_i\uparrow_j} = (d_1 - d_0)^2$. Thus, the effective molecular Hamiltonian is given by,
\ba
H =  g \sum_{i, j } \frac{1}{R^3} [ 1 - 3 \cos^2(\Phi - \Phi_0) \sin^2 \Theta_0] [- \frac{d_{01}^2 }{2}(a^\dagger_i a_j + a^\dagger_j a_i) + (d_1 - d_0)^2 n_i n_j ] 
\ea
Assuming that the electric field is pointed along $Z$ (perpendicular to the plane where the molecules are sitting), $\Theta_0 = 0$, meaning that the Hamiltonian simplifies to
\ba
H = g \sum_{i, j }\frac{1}{R^3}  [- \frac{d_{01}^2 }{2}(a^\dagger_i a_j + a^\dagger_j a_i) + (d_1 - d_0)^2 n_i n_j ] 
\ea
Thus, the natural Hamiltonian that emerges is a $1/R^3$, U(1) conserving hardcore bosonic Hamiltonian. Note that the ratio of the interactions to the hopping, $\frac{(d_1 - d_0)^2}{d_{01}^2}$ can be tuned by varying the strength of the electric field (see Fig.~2, maintext). In the limit, $|E| \rightarrow 0$, $(d_1 - d_0)^2 \rightarrow 0$, leaving only $H =  \frac{1}{R^3}  [- \frac{d_{01}^2 }{2}(a^\dagger_i a_j + a^\dagger_j a_i)$]. It is useful to rewrite the  Hamiltonian in terms of spin degrees of freedom:  $S^+_i = b^\dagger_i$, $S^-_i = b^\dagger_i$, $S^z_i = n_i -1/2$, wherein (upon dropping constant shifts) \cite{gorshkov2011quantum},
\ba
H = g \sum_{i, j }\frac{1}{R^3}  [- \frac{d_{01}^2 }{2} (S^+_iS^-_j + S^+_jS^-_i ) + (d_1 - d_0)^2  S^z_i S^z_j  ].
\ea

\vspace{1mm}

\noindent While interesting, the fact that the hopping term of the above Hamiltonian is unfrustrated implies that there are two natural phases. In the limit where the hopping dominates, we expect an easy-plane XY antiferromagnet, while in the limit where interactions dominate, we expect a crystal.

\bibliography{references}